# Linear instability of plane Couette and Poiseuille flows

Sergey G. Chefranov [1)], Alexander G. Chefranov [2)]

[1)] A. M. Obukhov Institute of Atmospheric Physics, Moscow, Russia;
e-mail: schefranov@mail.ru

[2)] Eastern Mediterranean University, Famagusta, Northern Cyprus;
e-mail: Alexander.chefranov@emu.edu.tr

## Abstract

It is shown that linear instability of plane Couette flow can take place even at finite Reynolds numbers Re > Reth ≈ 139, which agrees with the experimental value of Reth ≈ 150 ± 5 [16, 17]. This new result of the linear theory of hydrodynamic stability is obtained by abandoning traditional assumption of the longitudinal periodicity of disturbances in the flow direction. It is established that previous notions about linear stability of this flow at arbitrarily large Reynolds numbers relied directly upon the assumed separation of spatial variables of the field of disturbances and their longitudinal periodicity in the linear theory. By also abandoning these assumptions for plane Poiseuille flow, a new threshold Reynolds number Reth = 1035 is obtained, which agrees to within 4% with experiment—in contrast to 500% discrepancy for the previous estimate of Reth ≈ 5772 obtained in the framework of the linear theory under assumption of the "normal" shape of disturbances [2].

Показана возможность линейной неустойчивости плоского течения Куэтта (ПК) уже при числах Рейнольдса $Re > Re_{th} \approx 139$, что согласуется с полученной в эксперименте величиной $Re_{th} \approx 150 \pm 5$ (S. Bottin, et.al.,1997,1998). Этот новый результат линейной теории гидродинамической устойчивости получен на основе отказа от традиционно используемого предположения о продольной периодичности возмущений вдоль направления движения жидкости. Установленно, что ранее существовавшее представление о линейной устойчивости этого течения при любых сколь угодно больших числах Рейнольдса непосредственно связано именно с использованием в линейной теории предположения о разделении переменных пространственной изменчивости для поля возмущений и их продольной периодичности. При отказе от указанных предположений также и для плоского течения Пуазейля (ПП) получена новая величина порогового числа Рейнольдса $Re_{th} \approx 1035$, которая с точностью около 4% согласуется с экспериментом, в отличие от более чем 500% различия для ранее известной оценки $Re_{th} \approx 5772$, полученной также в рамках линейной теории, но при использовании " нормальной " формы для возмущений ( S. A. Orszag, 1971).

### 1.Introduction

Understanding the mechanisms of the loss of stability of a laminar regime, which results eventually in the transition to a new turbulent regime of flow in a medium, is currently a problem. Indeed, according to the linear theory of hydrodynamic stability for the Hagen–Poiseuille (HP) flow in a round tube, a transition from the laminar to any other flow regime cannot take place at any finite value of the threshold Reynolds number (Re th), which is at variance with experimental data. An analogous conclusion of the linear theory was also made and assumed to be valid until now for plane Couette (PC) flow [1]. For plane Poiseuille (PP) flow, well-known estimates of the threshold Reynolds number based on the linear theory were more than five times greater than the values observed in experiments [1–3].

As a result, the aforementioned transition between the laminar and turbulent regimes for HP, PC, and PP flows could be described only in the framework of the nonlinear theory of finite-amplitude disturbances [1]. However, the fact that this transition in HP, PC, and PP flows must always correspond to a rigid finite amplitude mechanism of the loss of stability, rather than to a soft mechanism described by the linear theory for disturbances of ultimately small amplitudes, is not yet established

In the present work, it is shown that linear instability of PC flow at finite Reynolds numbers Re > Reth is still possible, but only at the expense of abandoning traditional assumption of the "normal" representation of disturbances (according to which the field of disturbances exhibits periodic variations along the main flow direction and admits the separation of spatial variables). Previously, an analogous conclusion for HP flow was also obtained by abandoning the assumption about separation of spatial variables [4–6]. In the framework of a new theory of linear hydrodynamic stability of PP flow, a condition of the linear instability for this flow has been obtained which much better agrees with the quantitative experimental data than the estimate reported previously [2]. For PC flows, a condition of the linear instability obtained in this theory also agrees well with experimental data [7, 8], particularly for the threshold conditions of transition from the laminar to vortexed, yet nonturbulent, flow regime.

In the linear theory developed below, the possibility of quasi-periodic longitudinal (in the main flow direction) variations for which the longitudinal and transverse (orthogonal to solid boundaries) spatial variables in description of the field of disturbances cannot be separated. The energy method employed takes into account the existence of various periods of longitudinal variation for different transverse modes corresponding to the evolution of ultimately small ((perpendicular to both the flow direction and the normal to solid boundaries) velocity field disturbances and zero boundary conditions on the solid boundaries for PC and PP flows.

In the present work, it is shown (similar to [4–6] for the HP flow) that a possible mechanism of linear instability for PC and PP flows can be based on realization of dissipative instability. In turn, this mechanism of instability is characteristic of many physical systems where an important role is played by disturbances with negative energy [9–11]. This mechanism was previously related to the concept of "secular" instability [7]. This article consists of five sections. In Section 2, a new formulation of the problem of linear instability of PP and PC flows is given based on a new presentation of spatial variation of the field of disturbances (without traditional consideration of purely periodic fields with separable variables). In Section 3, the energy method is applied to PP and PC flows. In Section 4, the theoretical results are compared to experimental data, and in Section 5, final conclusions are formulated and discussed.

## 2. The statement of the linear instability problem for PP and PC flows

Let us consider conventional notions [1] about PP and PC flows of a viscous incompressible liquid bounded in the direction of the z axis by two solid parallel planes spaced by distance H. For PP flow (Fig. 1a) in the positive direction of the x axis, the origin of coordinates is set at the middle of the liquid layer (where the flow velocity is at a maximum) and the immobile boundaries have coordinates $z = H/2$ and $z = -H/2$. In case of PC flow (Fig. 1b), the origin is also set at the middle of the layer, but the liquid at $z=0$ is assumed to possess zero velocity. By analogy with [3], it is also assumed that the solid boundary at $z = -H/2$ moves at a velocity of $-V_{max}$, while the boundary at $z = H/2$ moves at a velocity of $+V_{max}$ in the positive direction of the x axis.

The linear stability of these flows is studied in a simple case where only ultimately small disturbances of the transverse (perpendicular to the flow direction and the normal to solid boundaries) velocity field component along the y axis take place. Let the fields of velocity and pressure disturbances be independent of coordinate y. Instead of the "normal" form of disturbances, we consider quasi-periodic disturbances, for which the spatial variables cannot be separated in description of the variation of disturbances depending on coordinates x and z (Fig. 1).

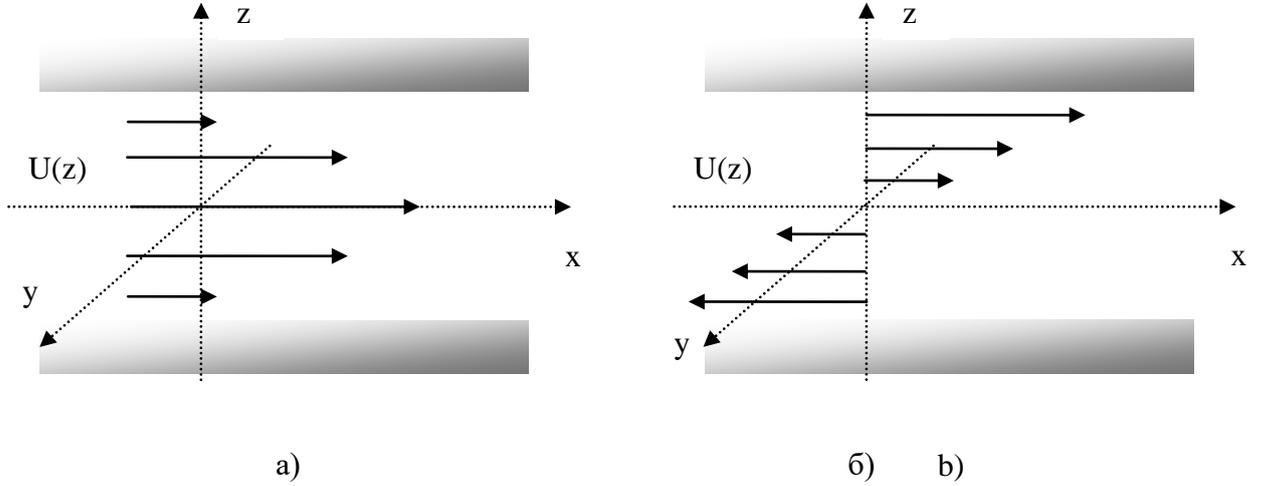

а)                                                          б)   b)

Fig. 1.Profiles of the main flow velocity U(z) normalized to Vmax for PP flows with (a) $U(z) = 1 - z^2$ and (b) $U(z) = z$ ) between solid boundaries at z = ±1 (in units of layer half-thickness H/2).

In a dimensionless form, the field of disturbances is described by the following equation and boundary conditions:

$$\frac{\partial V_y}{\partial \tau} + U(z)\operatorname{Re}\frac{\partial V_y}{\partial x} = \Delta V_y; \Delta = \frac{\partial^2}{\partial z^2} + \frac{\partial^2}{\partial x^2};$$
$$V_y(z = \pm 1) = 0,$$    (1)
$$U(z) = 1 - z^2, for PP; U(z) = z, for PC$$

Here, Vy is the dimensionless velocity (for the dimensional velocity of disturbances normalized to Vmax and the spatial coordinates are reduced to the dimensionless form using H/2), $\tau = 4t\nu/H^2$ - is the dimensionless temporal coordinate (t -is the dimensional time and ν -is the kinematic viscosity), and $\operatorname{Re} = \frac{V_{max}H}{2\nu}$ - is the Reynolds number for PP and PC flows (this definition coincides with that used in [2, 3]). In this problem setting, other components of the velocity field disturbances are absent (i.e., assumed to be zero for the sake of simplicity). It should be noted that a different transverse component Vz (rather than Vy) was considered in [2, 3]. This circumstance implied the necessity of solving a much more complex Orr–Sommerfeld equation [2, 3] instead of relatively simple equation (1). The latter statement is evident in view of the impossibility of excluding (without violation of zero boundary condition) the dependence of Vz on z and the need (following from the continuity equation) for considering at least one more nonzero component of the field of velocity disturbances.

Let us seek a solution of linear equation (1) in the following form that satisfies zero boundary condition in the direction of the z axis on solid boundaries:

$$V_y = e^{\lambda \tau}V; V = \sum_{n=0}^{N}(A_n(x)\sin(\pi z n) + B_n(x)\cos(\frac{\pi z(2n-1)}{2})); \lambda = \lambda_1 + i\lambda_2; V = V_1 + iV_2, V^* = V_1 - iV_2;$$
$$A_n(x) = A_n(x + T_n); B_n(x) = B_n(x + T_n); T_n = 1/\alpha_n; \max T_n = 1/\alpha_0; \alpha_0 < \alpha_1 < ... < \alpha_N; i^2 = -1$$
(2)

These relations in (2) determines new periodic boundary conditions along the x axis, which are set individually for each n-th mode *(n=0,1,2,...,N)*. The longitudinal periods Tn can be represented in a more general form as $T_n = q_n /(k_n \alpha_n)$, where $q_n, k_n$ - are an arbitrary natural numbers.

Thus, instead of the traditional "natural" representation of disturbances (with a single periodic boundary condition $T_n = T = const$ common to all modes), we introduce N periodic boundary conditions along the x axis, which are determined individually for each nth transverse mode. The new setting of the problem of linear theory of hydrodynamic stability in the form of Eqs. (1) and (2) for arbitrary longitudinal quasi-periodic disturbances differs from the previous formulation and can provide a better fit to experimental data (which reveal only a quasi-periodic rather than purely periodic spatial variation of the amplitude of disturbances in the direction of liquid motion between solid planes [1, 3]).

### 3. Energetic method

Let us consider on the basis of Eqs. (1), (2) the evoluition in time of the average (on the unit of mass) energy of disturbances:

$$E = \langle V_y V_y^* \rangle / 2 = e^{2\lambda_1 \tau} \langle VV^* \rangle / 2;$$
$$I_0 = \langle VV^* \rangle = \frac{1}{2}\int_{-1}^{1} dz \frac{1}{T_{max}} \int_0^{T_{max}} dx VV^* \quad (3)$$

Where in (3) $T_{max} = 1/\alpha_0$ -may be suggested.

### 3.1. Instability condition for PP flow

Using of the Eqs. (1)- (3), it is possible to obtain the following equation for exponent $\lambda_1$ that determines the increase (for $\lambda_1 > 0$) or decrease (for $\lambda_1 < 0$) in energy E with time for PP flow (for the sake of simplicity, in this case, we may set $B_n = 0$ in (2)):

$$2\lambda_1 I_0 = I_1 \text{Re} - I_2; I_2 = -\langle V^* \Delta V + V \Delta V^* \rangle > 0$$
$$I_1 = -\left\langle (U(z) \frac{\partial (VV^*)}{\partial x} \right\rangle = -\frac{1}{T_{max}} \sum_{n=1}^{N} \sum_{\substack{m=1 \\ m \neq n}}^{N} q_{nm} (A_n(T_{max}) A_m^*(T_{max}) - A_n(0) A_m^*(0)), \quad (4)$$
$$q_{nm} = \frac{1}{2}\int_{-1}^{1} dz(1-z^2)\sin(\pi z n)\sin(\pi z m) = \frac{4(-1)^{n+m+1} nm}{\pi^2 (n^2 - m^2)^2} \; if : n \neq m;$$

In order to derive Eq. (4), it is necessary to write, using Eq. (1) and the related equation for complex conjugate function $V_y^*$, the equation of energy evolution with allowance for definition (3) and then to substitute the solution in the adopted form (2).

For the sake of simplicity, let us consider Eqs. (2) – (4) with $A_n(x) = A_{0n} \exp(i 2\pi \alpha_n x)$. In this case, Eq. (4) yields**:**

$$I_2 = 2\sum_{n=1}^{N}(n^2 + 4\alpha_n^2)\pi^2 A_{0n}^2; I_0 = \sum_{n=1}^{N} A_{0n}^2; \quad (5)$$

$$I_1 = 2\alpha_1 \sum_{n=1}^{N} \sum_{\substack{m=1 \\ m \neq n}}^{N} q_{nm} A_{0n} A_{0m} \sin^2(\pi(p_n - p_m)) \ , \qquad (6)$$

where $p_n = \alpha_n / \alpha_1$. Let us $p_1 = 1; p_2 = \alpha_2 / \alpha_1 \equiv p; p_3 = \alpha_3 / \alpha_1 ...; p_N = \alpha_N / \alpha_1$.

In this representation for parameters $p$ and $p_n, n \geq 3$, the value of $\alpha_1$ is fixed and, for the sake of simplicity, we set it equal to unity. This corresponds to the period of disturbance of mode number 1, which coincides with the base longitudinal period (along the x axis) determined by half-thickness H/2 of the liquid layer. For example, the longitudinal period of disturbance for p = 2 corresponds to H/4, while that for p =0.5 corresponds to H. The convergence of the sum in expression (5) for $I_2$ in the limit as $N \to \infty$ is achieved with the following limitation on the initial amplitude:

$$A_{0n} \leq 1/n^{\frac{3+k}{2}}, k > 0.$$

From Eq. (4) it is possible to obtain the condition for linear instability of the PP flow:

$$\mathrm{Re} > \mathrm{Re}_{th} = I_2 / I_1 \qquad (7)$$

Let us use expressions (5) and (6) to minimize the right-hand side of the relation (7) with respect to parameter $\alpha_1$. Note that parameters $p_n$, defined above are assumed to be constant and independent of the $\alpha_1$ value. It can readily be checked that the right-hand side of relation (7) reaches a minimum at $\alpha_1 = \alpha_{1\min} = \sqrt{\frac{a}{b}}$. As a result, the minimum threshold Reynolds number with respect to parameter $\alpha_1$ is obtained:

$$\mathrm{Re}_{th\min} = \frac{\pi^4 \sqrt{ab}}{2c} \ , \qquad (8)$$

$$a = \sum_{n=1}^{N} \frac{1}{n^{1+k}}; b = 4 \sum_{n=1}^{N} \frac{p_n^2}{n^{3+k}}; c = -\sum_{n=1}^{N} \sum_{\substack{m=1 \\ m \neq n}}^{N} (-1)^{n+m} \frac{\sin^2(\pi(p_n - p_m))}{(n^2 - m^2)^2 (nm)^{\frac{1+k}{2}}} \ . \qquad (9)$$

In the general case, the value of (8) should be minimized with respect to the free, continuously variable parameters $k, p, p_3, ..., p_N$.

For the sake of simplicity, let us restrict consideration to the case where the minimization of expression (8) is performed only for parameters $k$ and $p$, while the other parameters are assumed to be fixed and sufficiently slowly increasing functions of the corresponding number $n=3,4, ...,N$ (for the convergence of series (9) for $b$, let $p_n = n^{\frac{k}{8}}$ for $n \geq 3$). Figure 2a shows the results of calculations using Eqs. (8), (9) for PP flow. The minimum threshold Reynolds number in this case, $\mathrm{Re}_{th\min} = 1035.3$, is achieved at N=100 for k=0.675 и p=0.506.

### 3.2 Instability condition for PC flow

Let us consider Eq. (1) in case of PC flow. It is important to note that $B_n \neq 0$ in the (2). For example, in the case $A_n = B_n$ it is possible to obtain from (2) and (3) in (1):

$$I_0 = 2\sum_{n=0}^{N} A_{0n}^2, I_2 = 2\pi^2 \sum_{n=0}^{N} A_{0n}^2 (4\alpha_0^2 p_n^2 + \frac{1}{2}(n^2 + \frac{(2n-1)^2}{4})), p_n = \frac{\alpha_n}{\alpha_0}, p_0 = 1, p_1 = p, p_n = n^{\frac{k}{8}}, n > 1$$

$$I_1 = 8\alpha_0 \sum_{\substack{n=0 \\ m \neq n}}^{N} \sum_{m=0}^{N} \frac{(-1)^{n+m} A_{0n} A_{0m} \sin^2(\pi(p_n - p_m))}{\pi^2} \left( \frac{n(m-\frac{1}{2})}{(n^2 - (m-\frac{1}{2})^2)^2} + \frac{m(n-\frac{1}{2})}{(m^2 - (n-\frac{1}{2})^2)^2} \right)$$

(10)

From (7) and (10) the condition of the PC linear instability is obtained.

Let us use expressions (10) to minimize the right-hand side of the relation (7) with respect to parameter $\alpha_0$ when for the case $A_{0n} = \dfrac{A_0}{(1+n)^{\frac{3+k}{2}}}$ the threshold value (8) is obtained. Now for the case of PC flow in (8) we must use the relations:

$$a = \frac{1}{2}\sum_{n=0}^{N} \frac{(n^2 + \frac{(2n-1)^2}{4})}{(1+n)^{3+k}}, \quad b = 4\sum_{n=0}^{N} \frac{p_n^2}{(1+n)^{3+k}},$$

$$c = \sum_{\substack{n=0 \\ m \neq n}}^{N} \sum_{m=0}^{N} \frac{(-1)^{n+m}}{[(1+n)(1+m)]^{\frac{3+k}{2}}} \sin^2(p_n - p_m) \left[ \frac{n(m-\frac{1}{2})}{(n^2 - (m-\frac{1}{2})^2)^2} + \frac{m(n-\frac{1}{2})}{(m^2 - (n-\frac{1}{2})^2)^2} \right].$$

(11)

Figure 2b shows the results of calculations using Eqs. (8) and (11) for PC flow. The minimum threshold Reynolds number in this case, Reth min = 139.077, is achieved at N = 100 for k = 1.7037 and p = 0.4859.

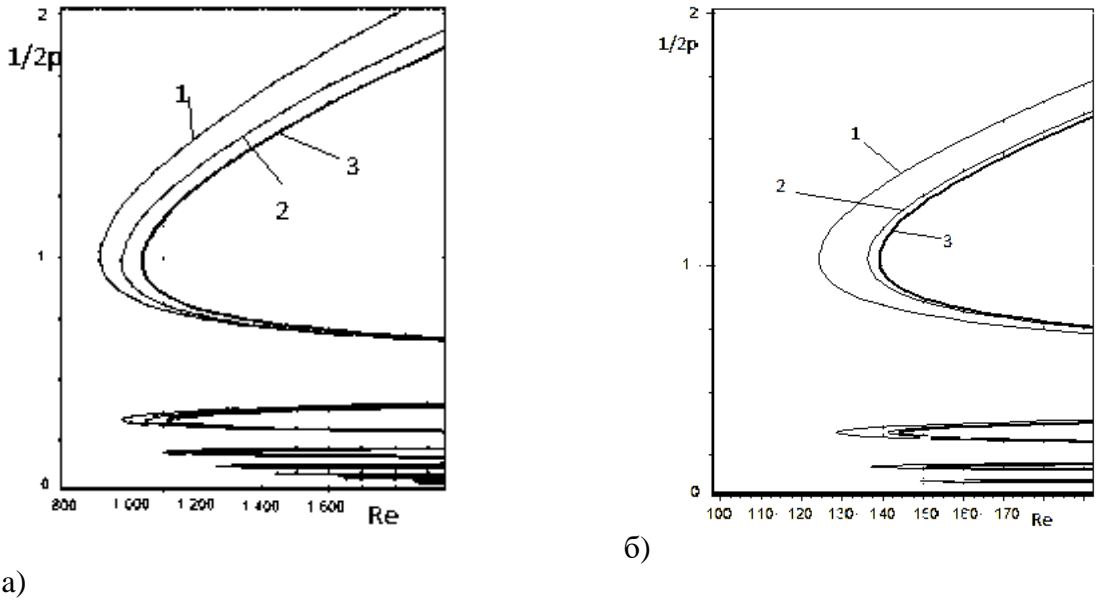

a)                                                                            б)

Fig. 2 Curves of neutral stability of the (a) PP flow for k = 0.675 and (b) PC flow for k = 1.7037:
(1) N = 2, Reth min = 906.35, 1/2p = 1.008; (2) N = 10, Reth min = 972.825, 1/2p = 0.988; (3) N = 100, Reth min = 1035.3, 1/2p = 0.988; (4) N = 2, Reth min =124.273, 1/2p = 1.029; (5) N = 10, Reth min = 136.475, 1/2p = 1.029; (6) N = 100, Reth min = 139.077, 1/2p = 1.029.

In Fig. 2a, fragments of the neutral stability curve corresponding to conditions (7)–(9) are plotted in a scaled form as 1/2p versus Re for PP flow; in Fig. 2b, the analogous curves are presented for PC flow in terms of the linear instability criterion (7) and (8) with allowance for expressions (10) and (11).

It should be emphasized that, since only the transverse component of disturbance of the velocity field is considered in Eq. (1), the mass flux for superposition of the main flow and disturbance field is retained.

Although spontaneous appearance of such disturbances is naturally not excluded, it becomes highly improbable with increasing smoothness of the boundary surface. On the other hand, these disturbances can be artificially created in laboratory modeling of PP and PC flows [7, 8].

## 4. COMPARISON TO EXPERIMENTAL DATA

### 4.1 PP flow

Early experimental results for PP flow were reported, e.g., in [12], where (unlike the present work and [3]) the Reynolds number $\text{Re}_D$ was defined via the cross-section-averaged flow velocity $V_a = \frac{2V_{\max}}{3}$ and total layer thickness H. The minimum threshold Reynolds number in [12] was found to be

$$\text{Re}_D = \frac{V_a H}{\nu} = \text{Re}_{Dth} \approx 1440.$$

The relationship of $\text{Re}_D$ to Reynolds number Re as defined in the present work is $\text{Re}_D = \frac{4\text{Re}}{3}$. Taking this into account, we conclude that the stability threshold for PP flow obtained in [12] corresponds to $\text{Re} > \text{Re}_{th} \approx 1080$, which differs by less than 4% from the estimate of $\text{Re}_{th} \approx 1035$ obtained in the present work.

### 4.2 PC flow

Experimental results for PC flow were reported in [7, 8], where the flow stability was studied using artificial disturbances generated by a thin wire of radius ρ (0.0036 < ρ/h < 0.1714, h = H/2). The wire was stretched along the y axis (see Fig. 1b), so that the resulting disturbances evidently had a nonzero $V_y$ component of the flow velocity disturbance field, the evolution of which is studied in the present work based on Eq. (1). Thus, PC flow in [7, 8] exactly corresponded to the scheme in Fig. 1b and was rather insignificantly modified by the presence of the wire.

The Reynolds number defined in experiments [7, 8] coincided with that used in the present work, and the accuracy of determining the threshold Reynolds number in experiments varied from 2.5 to 4%.

Data reported in [7, Fig. 2] and [8, Fig. 12] were presented as diagrams of the transition from basic laminar PC flow to a vortexed but relatively regular (nonturbulent) regime with a Reynolds number above the first threshold value $\text{Re} > R_0(\rho/h)$ dependent on the wire thickness (Fig. 3).

For the comparison of results of the present work to experimental data [7, 8], Fig. 3 reproduces the plot of [8, Fig. 12] with a superimposed fragment of the neutral stability curve 6 from Fig.2b. The region situated above curve 1 in Fig. 3 corresponds to the linear (exponential) instability of the PC flow.

Thus, the region of linear instability of the PC flow bounded by curve 1 is very close to (sometimes, even coinciding to within experimental accuracy with) the experimental curve describing the boundary of transition at $\text{Re} > R_0$ from the observed laminar to vortexed regime preceding the appearance of turbulent regions (spots) in the flow. These turbulent spots were observed already at $Re > R_2 \approx 325$ [7, 8].

Thus, Figs. 2b and 3 show that the threshold Reynolds number is sensitive to changes in parameter p. For example, the threshold Reynolds number at $1/2p = 0.5$ is infinitely large, but decreases to $Re_{th} \approx 305$ already at $1/2p = 0.67$ and to $Re_{th} \approx 139$ at $p = 1/2p = 1.029$.

This behavior characterizes the important role of parameter p related to frequency–wave properties of the field of disturbances, independent of their amplitude. Note that the sensitivity of the threshold Reynolds number to the wire thickness observed in experiments [7, 8] also characterizes the dependence of the R0 value on the ratio of wavelengths of initial disturbances rather than of their amplitudes. It should also be noted that the range of variation of parameter 1/2p in Fig. 3 corresponds to a decrease in this value by a factor of 1.536 (from 1.029 to 0.67). Indeed, a change in the wavelength of vortex disturbances by a factor of 1.375 (i.e., close to 1.5) was observed [8, Fig. 13a] upon a change in the ρ/h value from 0.01 to 0.085. For this reason, the experimental curve in Fig. 3 (and in [8, Fig. 12]) would remain the same upon replacing parameter ρ/h on the axis of abscissas by the corresponding disturbance wavelength expressed (like 1/2pvalue in this work) in units of h = H/2. Note also the possible relationship between the characteristic longitudinal periods of initial disturbances and observed periods (along the y axis in Fig. 1b) of vortex disturbances for Re > R0 in [7, 8].

The data reproduced from [7, 8] (see Fig. 3, where R = Re) show that the second threshold Reynolds number R2 ≈ 325 (characterizing the transition to turbulence at Re > R2) remains almost unchanged when parameter ρ/h varies from 0.02 to 0.1, whereas R2 increases to about 375 when ρ/h decreases from 0.02 to 0.005. Evidently, the transition from the laminar to turbulent regime must be related to nonlinear processes that are not considered in the present work. It should be noted that, after investigations in [7, 8], the subsequent experiments were probably not aimed at elucidating mechanisms of the transition from basic laminar PC flow to the vortexed (nonturbulent) regime of flow at Re ≥ R0. The main purpose of current experiments is still the elucidation of specific features of the coexistence of turbulent spots and laminar regions in PC flow at Reynolds numbers above R2 [13].

In this respect, results of the present work can stimulate experimental investigations aimed at filling the gap particularly in the transition from a laminar to non laminar flow regime.

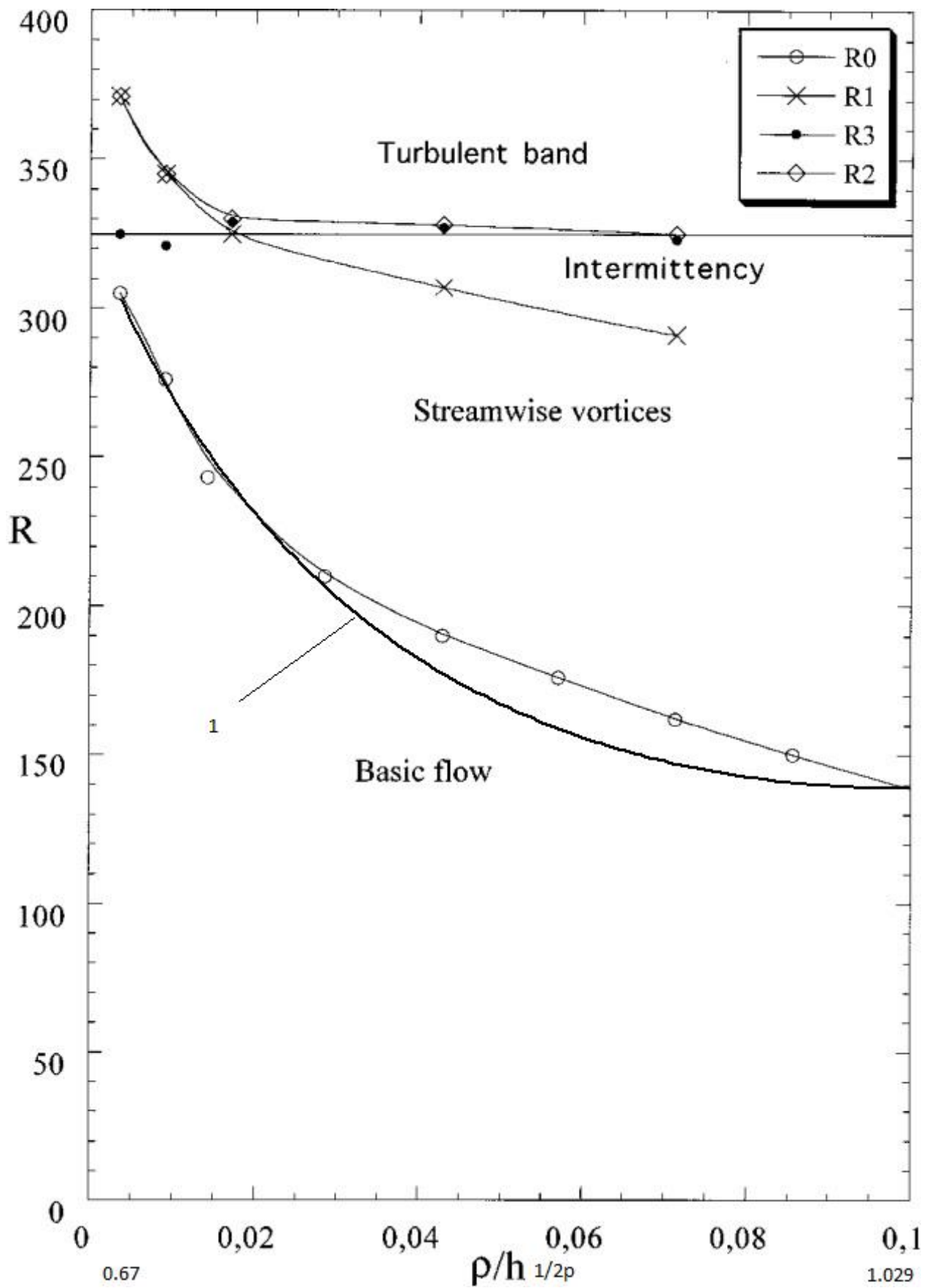

Fig.3 Comparison of published experimental data [8, Fig. 12] for PC flow with (curve 1) results of the present work (fragment of curve 6 from Fig. 2b); 1/2p value varies along the axis of abscissas from 0.67 (corresponding to ρ/h = 0.005) to 1.029 (ρ/h = 0.1).

## 5. Discussion and conclusions

The conclusion concerning a possible linear instability of PC and PP flows was derived in this work from Eq. (1), describing evolution of the transverse component of the velocity field disturbances, under condition that the right-hand side is nonzero owing to a finite value of the coefficient of kinematic viscosity.

Here, an important condition for the possible realization of instability is not the right-hand part of Eq. (1) as such (and the corresponding integral I2 in Eq. (4)), but a zero boundary condition that follows from it at any value of ν > 0 and is taken into account in representation of the disturbance field in the form of Eq. (2).

Previously, an analogous mechanism of hydrodynamic dissipative instability was considered by L. Prandtl in 1921–1922 in studying the stability of a laminar boundary layer and by W. Heisenberg (1924) and S.S. Lin (1944–1945) in establishing linear instability of PP flow. More recently, an example of the dissipative instability of a linear two-dimensional oscillator (with linear friction with respect to velocity) in a rotating coordinate system was given in [10], which was previously considered by Lamb [9] as an example of secular instability.

It was additionally established in [10] (in contrast to [9], where attention was not paid to this issue) that the indicated centrifugal dissipative instability is related to violation of the chiral symmetry determining the mechanism of cyclone–anticyclone asymmetry in the atmospheres of rapidly rotating planets. Understanding of the phenomenon of dissipative instability of flows near solid boundary surfaces can also be achieved with the aid of the Landau method [19], which was originally used for estimating the critical velocity of a superfluid moving in a capillary on the basis of analysis of vortex disturbances (rotons) possessing negative energy.

The conclusions obtained above and in [4–6] (where conditions of linear instability of HP flow in a tube with round cross section were established) fill the well-known gap in the nonlinear theory [1], where only a stage of algebraic instability is considered instead of seed linear (exponential) instability.

It should also be noted that it would be of interest to consider the linear instability established for PC flow in application, e.g., to the problem of appearance of so-called killer waves in the ocean, which are frequently observed in regions with rather strong streams (Kurosio, Gulf Stream, etc.) characterized by relatively high shear velocities [14].

The mechanisms proposed previously did not consider the possible excitation of these waves as related to the hydrodynamic instability of the corresponding shear flows with respect to disturbances of extremely small amplitude.

According to the theory considered above, these disturbances of the velocity field may have a component transverse to both the horizontal flow velocity direction and the horizontal direction in which this velocity changes. It is clear that component $V_y$ of the velocity disturbance component of the PC flow in Eq. (1) can also describe vertical motions which, on reaching the water surface, can lead to the appearance of waves with anomalously large amplitudes. Although the well-known mechanism of modulation instability [14–16] is based on the linear theory, it nevertheless contains a threshold condition for the disturbance amplitude according to which an instability develops provided that

$$A_0 > A_{0th} = \Delta\Omega / \sqrt{2}\omega_0 k_0$$

where $A_0, k_0, \omega_0, \Delta\Omega$ - are the initial amplitude, wavenumber, disturbance wave frequency, and modulation frequency (in [14], this condition was alternatively written as $\varepsilon N > 1/\sqrt{2}, \varepsilon = k_0 A_0, N = \omega_0 / \Delta\Omega$).

It is important to point out that the mechanism related to the linear instability of PC flow considered in this work involved no limitations on the amplitude of disturbances.

In future investigations, it would be of interest to consider application of the energy approach developed above to estimation of the minimum threshold Reynolds number in circular Couette flow, for which the problem of linear stability is absent (in contrast to the case of PC and PP flows).

In addition, it would be desirable to carry out direct numerical calculations of the stability of PC and PP flows (analogous to those performed in [17, 18]) for refining the boundaries of applicability of the energy method.

ACKNOWLEDGMENTS

This work was supported in part by the Russian Science Foundation, project no. 14-17-00806.

**REFERENCES**

1. A. S. Monin and A. M. Yaglom, Statistical Fluid Mechanics: Mechanics of Turbulence (Gidrometeoizdat, St. Petersburg, 1992; MIT, Cambridge, MA, 1971), Vol. 1.

2. S. A. Orszag, J. Fluid Mech. 50, 689 (1971). Doi 10.1017/S0022112071002842

3. S. A. Orszag and L. C. Kells, J. Fluid Mech. 96, 159 (1980). doi 10.1017/S0022112080002066

4. S. G. Chefranov and A. G. Chefranov, arXiv: 1007.1097v1 [physics.flu-dyn].

5. S. G. Chefranov and A. G. Chefranov, J. Exp. Theor. Phys. 119, 331 (2014). doi 10.1134/S1063776114070127

6. S. G. Chefranov and A. G. Chefranov, Dokl. Phys. 60, 327 (2015). doi 10.1134/S1028335815070083

7. S. Bottin, O. Dauchot, and F. Daviaud, Phys. Rev. Lett. 79, 4377 (1997). doi 10.1103/PhysRevLett.79.4377

8. S. Bottin, O. Dauchot, F. Daviaud, and P. Manneville, Phys. Fluids 10, 2597 (1998). doi 10.1063/1.869773

9. H. Lamb, Hydrodynamics (Cambridge Univ. Press, Cambridge, 1993)

10. S. G. Chefranov, JETP Lett. 73, 274 (2001). doi 10.1134/1.1374259

11. S. G. Chefranov, Phys. Rev. Lett. 93, 254801 (2004). doi 10.1103/PhysRevLett.93.254801

12. S. J. Davies and C. M. White, Proc. R. Soc. London A 119, 92 (1928). doi 10.1098/rspa.1928.0086

13. M. Couliou and R. Monchaux, Phys. Fluids 27, 034101 (2015). doi 10.1063/1.4914082

14. M. Onorato and D. Proment, Phys. Rev. Lett. 107, 184502 (2011). doi 10.1103/PhysRevLett.107.184502

15. V. Zakharov, J. Appl. Mech. Tech. Phys. 9, 190 (1968). doi 10.1007/BF00913182

16. T. B. Benjamin and J. E. Feir, J. Fluid Mech. 27, 417 (1967). doi 10.1017/S002211206700045X

17. J. Rolland, arXiv: 1401.3586v1 [physics.flu-dyn].

18. S. Zammert and B. Eckhardt, arXiv: 1312.6783v1 [physics.flu-dyn]. 19. L. D. Landau, Zh. Eksp. Teor. Fiz. 11, 592 (1941)